\documentstyle[aps,prl,multicol,epsf]{revtex}
\begin{document}

\title{Denaturation of Heterogeneous DNA}
\author{Dinko Cule and Terence Hwa}
\address{Department of Physics, University of California at San Diego, La Jolla, CA
92093-0319}
\date{revised: \today }
\maketitle

\begin{abstract}
The effect of heterogeneous sequence composition on the denaturation of 
double stranded DNA is investigated. The resulting pair-binding energy
variation is found to have a negligible effect on the critical properties
of the smooth second order melting transition in the simplest 
(Peyrard-Bishop) model. However, sequence heterogeneity is dramatically
amplified upon adopting a more realistic treatment of the backbone stiffness.
The model yields features of ``multi-step melting'' similar to those observed
in experiments.
\vspace{10pt}
\end{abstract}
\begin{multicols}{2}
\narrowtext

The denaturation or melting of double-stranded DNA molecules 
upon changes in ambient
temperatures or solvent conditions is a subject which has had a long history~
\cite{rev,rev1}. In early theoretical studies~\cite{rev,zimm}, DNA melting
was described by the nearest-neighbor 1d Ising model which yielded only a
sharp {\em crossover} but not a thermodynamic phase transition~\cite{landau}
. A smooth second order transition was later demonstrated in a modified
version of the 1d Ising model~\cite{long}, after including an effective {\em 
long-range} interaction arising from the configurational entropy gain of the
denatured segments. Experimentally, a purified DNA sample containing an
unique sequence and length is found to exhibit distinct multi-step melting,
where the ``melting curves'' (to be specified below) exhibit sharp features
consisted of plateaus of variable sizes separated by steps~\cite{rev1,gotoh}
. These fine features have been attributed to the melting of individual
``domains'' associated with variations in the composition of the nucleotide
sequences, since the binding energies of the two kinds of base pairs,
adenine-thymine (AT) and guanine-cytosine (GC), are significantly different~
\cite{bp}. The effect of binding energy variation has been studied
previously using the nearest-neighbor 1d Ising model~\cite{rev,montroll}. In
this paper, we investigate this effect in detail by incorporating the
important configurational entropy of the denatured strands in a systematic
way. We find that sequence heterogeneity itself is not sufficient to produce
multi-step melting, with the phase transition remaining to be of the
second-order. Nevertheless, heterogeneity can be dramatically amplified by
small changes in the detailed form of the configurational entropy, leading
to the apparent multi-step melting behavior for finite length sequences. 

A simple way of incorporating the entropy of the single-stranded segments is
to model the two single strands ${\bf r}_n^{(1)}$ and ${\bf r}_n^{(2)}$ by
random walks~\cite{sa}, with the index $n$ denoting the $n^{th}$ base pair.
The binding of complementary base pairs is described by a potential
function $V_n( {\bf r}_n^{(1)}-{\bf r}_n^{(2)}) $, where $V_n$ has a hard
core (reflecting the repulsion of the phosphate backbone) and an attractive
(short-ranged) tail mimicking the hydrogen bond between each base pair. 
Further taking into account the directional specificity of the hydrogen
bonds, one obtains the following Hamiltonian for a double strand of $N$ base
pairs, 
\begin{equation}
\beta {\cal H}=\frac 1T\sum_{n=1}^N\left\{ \frac K2\left( y_{n+1}-y_n\right)
^2+V_n(y_n)\right\} .  \label{H}
\end{equation}
In Eq.~(\ref{H}), $y_n$ is the component of the relative displacement field $%
{\bf r}_n^{(1)}-{\bf r}_n^{(2)}$ along the direction of hydrogen bond and is
the important degree of freedom we will focus on. The quadratic coupling
describes the stiffness of the backbone.

The model (\ref{H}) without the $n$-dependence in the interaction $V$ is
known as the Peyrard-Bishop (PB) model~\cite{pb}. Although PB specifically
used the Morse potential for $V$, the qualitative behavior of the system is
well known for a large class of short-ranged potentials, via a mapping to a
fictitious quantum mechanics problem~\cite{pg1}. Let the depth of the
(asymmetric) ``potential well'' $V(y)$ be $U_0$ and the range of attraction
be $a$, then a continuous phase transition\cite{transition} occurs at a
critical temperature $T_m$, given by the condition $T_m^2/(2Ka^2)\sim U_0$.
This phase transition is characterized by a discontinuity in the specific
heat $C$ and an algebraic divergence in the average separation distance $%
\ell $ between the base pairs, $\ell \sim (T_m-T)^{-\nu }$ with $\nu =1$.
The inverse of the pair distance $\ell ^{-1}$ is the order parameter of this
transition and can be directly related (see below) to the fraction of
unbroken base pairs, a key experimental observable~\cite{rev1}.

Next, we describe the effect of a variable interaction $V_n(y)$ which is
fixed by the DNA sequence. For simplicity, we restrict $y$ to the positive
real axis, and model the interaction as a product of a short ranged function 
$\delta _a(y)$ [with $\delta _a(0<y\lesssim a)\approx 1$ and $\delta
_a(y>a)\rightarrow 0$] and a variable strength $V_n$~\cite{note}. 
We assume the sequence to be
random and short-range correlated, with an average pairing energy $\overline{%
V_n}=-U_0$, and the fluctuation $\delta V_n=V_n-\overline{V}$ described by
its variance $\overline{\delta V_m\delta V_n}=\Delta \,\delta _{m,n}$.
(Here, the overbar denotes average over the ensemble of random sequences.)
Some important conceptual issues to understand are whether the melting
transition survives in the presence of the quenched-in variable interaction;
and if so, what is the nature of the transition.

These issues have been studied in the past decade in the context of some
closely related systems, e.g., one describing the adsorption of a Gaussian
random heteropolymer by a solid surface~\cite{shak,obukhov,joanny,chak,li}
(with $y_n$ being the distance of the $n^{th}$ monomer from the surface and $%
V_n(y)$ giving the interaction of that monomer with the surface), and
another describing the wetting of a 1d interface from a random substrate~%
\cite{orland,derrida,bulk}. It is known that the melting transition still
exists and the effect of quench-in randomness is {\em marginal} in the
renormalization group sense. This conclusion is straightforwardly reached,
for instance, by considering the perturbative effect of a weak disorder on
the melting temperature: Assuming that the effect of randomness is
negligible in the small $\Delta$ limit, then fluctuation in base separation $%
y_n$ is correlated along the backbone up to the length $\xi \sim \ell^2$. At
a temperature $T$ slightly below the melting temperature $T_m$ of the pure
system, $\xi \sim T_m^2/(T_m-T)^2$ becomes very long. Variation in the
interaction energy $U_0$ averaged over the scale $\xi$ is $\delta U \sim 
\sqrt{\Delta \xi}/\xi \approx \sqrt{\Delta} \, (T_m-T)/T_m$, which leads to
a {\em shift} in the melting temperature of the order 
\begin{equation}
\delta T_m \propto (T_m-T) \sqrt{\Delta}/U_0.  \label{dT}
\end{equation}
The effect of randomness is revealed by comparing $\delta T_m$ with $\delta
T \equiv T_m-T$ in the limit $\delta T \to 0$. Randomness is irrelevant if $%
\delta T_m \ll \delta T$, but is non-negligible if otherwise. 
The problem at hand is ``marginal" since $\delta T_m \sim \delta T$.

Much effort has been devoted to resolving whether the randomness is
marginally relevant or marginally irrelevant. Early studies~\cite
{shak,orland} suggest that it is marginally irrelevant, such that critical
properties of the melting transition are the same as those of the pure case
(up to logarithmic corrections). However, more recent renormalization-group
studies~\cite{derrida,bulk} find the randomness to be marginally relevant,
indicating that the scaling properties should be different from the pure
case beyond a crossover length $\xi _{\times }\sim \exp [c\,U_0^2/\Delta
]$ with $c~\sim O(1) $, or equivalently if the reduced
temperature is within the Ginzburg temperature $\delta T_{\times }\sim \xi
_{\times }^{-1/2}$. The actual scaling behavior in the asymptotic ``strong
coupling'' regime is however not accessible from these studies.

The theoretical results have so far not been carefully tested numerically:
In Ref.~\cite{orland}, evidence in support of the irrelevancy of randomness
was reported, while the contrary was claimed in Ref.~\cite{derrida}. The
numerics in the latter work was in fact rather qualitative; no attempt in
characterizing the alleged strong coupling regime was made. The numerics in
Ref.~\cite{orland} was flawed on the other hand by assuming a particular
value for the melting temperature which was later shown to be incorrect~\cite
{derrida}. We have thus re-investigated the nature of the melting transition
numerically.

As in Refs.~\cite{orland} and \cite{derrida}, we use a transfer matrix
calculation~\cite{tm}. We start with the transfer integral solution of the
``wave function" $\phi_n(y)$~\cite{pb}, 
\begin{equation}
\phi _{n+1}(y)=\int_{y^{\prime }}\,\exp \left[ -\frac K{2T}(y-y^{\prime })^2-%
\frac{V_n(y)}T\right] \,\phi _n(y^{\prime }).  \label{TI}
\end{equation}
To speed up the numerics, we placed the recursion relation (\ref{TI}) on a
lattice such that $y$ takes on only non-negative integer values, from $0$ to 
$L$. We further restricted $y_{n+1}-y_n\in \left\{ 0,\pm 1\right\} $. The
potential function $V_n(y)$ has the simple form $V_n(y=0)=-U_0\pm \sqrt{%
\Delta }$ with equal probability, and $V_n(y>0)=0$. To characterize the
system, we compute (for each configuration of the random potential $V_n$)
the ``steady-state'' probability distribution $P_n(y)$ of finding $y_n$ at $y
$. This is obtained (up to normalization) as the product of the forward- and
backward- propagated wave functions (see also \cite{derrida}). Sufficiently
long segments ($\sim 10^4$) close to the two ends of the sequence are
truncated to ensure that the results are independent of the choice of
boundary conditions. From the steady-state distribution, we compute the
moments $\left\langle y_n^m\right\rangle _L\equiv \sum_{y=0}^L\,y^m\,P_n(y)$%
, and then average over different realization of the randomness (or average
over $n$ for very long sequences). For example, the average pair distance $%
\ell $ is given by $\ell =\overline{\left\langle y_n\right\rangle }_L$.

We first identify the critical point by monitoring the $L$-dependence of the
dimensionless variance, $\delta y=\overline{[\left\langle y_n^2\right\rangle
_L-\left\langle y_n\right\rangle _L^2]^{1/2}}/\ell $, which should be $L$%
-independent at the critical point. This is shown in the inset of Fig.~1(a)
for systems with $U_0=-1,$ $\Delta =1$, $K=1$ and $N=10^5$, averaged over $10
$ independent realization of random sequences. The parameters are chosen
such that $\xi _{\times }\sim O(1)$ and the system is readily in the strong
coupling regime. [Note also that the unit of $n$ in our simplified discrete
model is no longer one base pair. With the above parameters, the length of
our system corresponds to a sequence of several thousand base
pairs of the PB model~\cite{pb}, with $\pm 20\%$ pair-to-pair variation in
the binding energy.] Using the empirically obtained value of $T_m$,  we plot
the pair distance $\ell $ vs.~the reduced temperature in Fig.~1(a). An
exponent $\nu =1$ is obtained, indicating the absence of anomalous scaling.
To further test the relevancy of the
\begin{figure}
\epsfysize=1.5in
\centerline{\epsffile{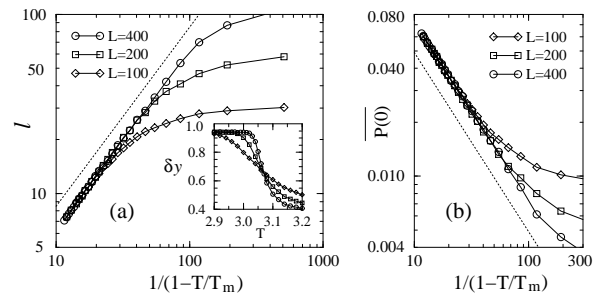}}
\vspace{\baselineskip}
\caption{(a) The average pair distance $\ell $ for various transverse system
sizes. The melting temperature $T_m\approx 3.07 $ is determined from the
plot in the inset. (b) The ``melting curves'', showing the fraction of
unbroken pairs $\overline{P(0)}$. Scaling laws with $\nu =1$ 
are shown by the dotted straight lines. }
\label{F1}
\end{figure}
\noindent randomness, we directly measure the
average fraction of unbroken pairs, $\overline{P_n(0)}$, which should scale
as $1/\ell \sim (T_m-T)$ if the randomness is irrelevant. However, as
pointed out in Ref.~\cite{lassig1}, relevancy of the randomness would imply
additional singularity in the distribution $P_n(y\rightarrow 0)$, resulting
in the anomalous scaling of $\overline{P_n(0)}$. This quantity is plotted
against the reduced temperature in Fig.~1(b). Again, we find no evidence
of anomalous scaling. Since our calculations are performed in
the strong-coupling regime, we conclude that either the randomness is
irrelevant or the asymptotic scaling in the strong-coupling regime is almost
indistinguishable from the pure problem. Similar results have been obtained
for a variety of different parameter choices.

While the ultimate resolution to the issue of the relevancy of sequence
heterogeneity may require yet larger systems with good statistics, 
it is clear from Fig.~1 that the
melting transition encountered here is rather smooth.
Indeed, the numerically obtained melting curve for a {\em single} sample is
very smooth (see Fig.~3(b) below), without any noticeable fine structures.
The smoothness of the transition is irrefutable even in the specific heat
curves used to support the relevancy of randomness in Ref.~\cite{derrida}.
This makes the experimentally observed multi-step melting behavior rather
puzzling. To investigate the possible cause of multi-step melting, we
recall a recent numerical finding~\cite{dpb} that fluctuation effect in DNA
melting is much enhanced upon adopting a more realistic form of the
backbone stiffness, to reflect the fact that the 
the DNA is significantly more rigid in the double-stranded conformation. 
An explicitly $y$ -dependent stiffness 
\begin{equation}
K(y_n,y_{n+1})=K_1+(K_2-K_1)\,e^{-(y_n+y_{n+1})/2b}  \label{K}
\end{equation}
was used in Ref.~\cite{dpb} to match the stiffness of the double strand $K_2$
(for $y=0$) and the single strand $K_1<K_2$ (for $y\rightarrow \infty $).
Numerical solution~\cite{dpb} of the homogeneous version of (\ref{TI}) using
the modified stiffness $K(y^{\prime },y)$ yields what appears to be a {\em %
first-order} melting transition!

Let us examine the effect of the variable stiffness in some detail. To
facilitate the analysis, we modify the exponential factor in (\ref{K}) to $%
e^{-y_n/b}$. This does not cause any significant differences since $%
y_n\approx y_{n+1}$. With the modified form of $K(y)$, one can
straightforwardly perform the transfer integral (\ref{TI}), which in the
continuum limit yields a Schr\"{o}dinger-like equation for $\phi_n(y)$,
with an effective potential $\widetilde{V}_n(y)=V_n(y)+\frac T2\log 
\left[ K(y)/K_1\right] +{\rm const}$, as well as an effective diffusion 
coefficient $T/2K(y)$. Let us
focus on the form of $\widetilde{V}$: In addition to the original attractive
potential $V_n(y)$ of range $a$, there is now a {\em repulsive} term of the
order $U_1=(T/2)\log (K_2/K_1)>0$ of range $b$. The latter plays the role of
an (entropic)  {\em barrier} and is the main effect introduced by the
variable stiffness 
$K(y)$. Qualitative features of the homogeneous system can be
obtained by solving the Schr\"{o}dinger equation using a constant
stiffness $K$ and a toy potential function $\widetilde{V}(y)$ 
on  the half space $y>0$, with $\widetilde{V}=-U_0+U_1$ for $y<a$, 

\begin{figure}
\epsfysize=1.45in
\centerline{\epsffile{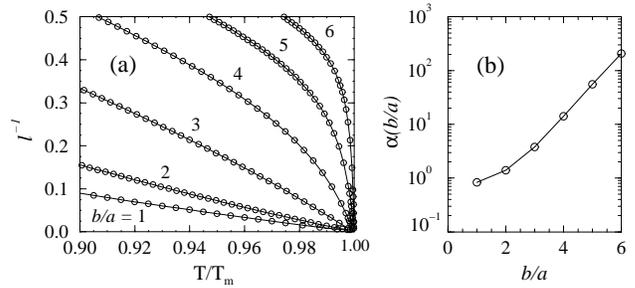}}
\vspace{\baselineskip}
\caption{ (a) Solution of the homogeneous PB model with a variable backbone
stiffness. The order parameter $\ell ^{-1}$ still vanishes linearly with the
reduced temperature. (b) The slope $\alpha $ has an exponential dependence on
the ratio $b/a$. $T_m$ itself changed by $\sim 50\%$ over the range of $b/a$
studied. 
}
\label{F2}
\end{figure}
\noindent 
$\widetilde{V}=+U_1$ for $a<y<b$, and 
$\widetilde{V}=0$ for $y>b$. We find the order parameter $\ell ^{-1}$ of the
homogeneous problem still to have the form $\ell ^{-1}=\alpha \cdot
(T_m-T)/T_m$ near $T_m$, although the amplitude $\alpha $ increases {\em %
exponentially} for increasing $U_1$ or $b/a$. Our finding is verified
numerically (Figs.~2) by exactly diagonalizing the transfer integral (\ref
{TI}) for the homogeneous problem, using the Morse potential for $V(y)$ and
Eq.~(\ref{K}) for $K(y,y^{\prime })$ as in Ref.~\cite{dpb}. Note that the
width of the transition region scales as $1/\alpha $. Thus, for the
parameter value $K_2/K_1=1.5$ and $b/a \approx 5$ used in Ref.~\cite{dpb}
(corresponding to $\alpha \approx 100$ according to Fig.~2(b)), the
transition region is extremely narrow, making it very much first-order-like
in appearance.

The heterogeneous system is studied next using the lattice transfer matrix
algorithm, incorporating the toy potential $\widetilde{V}$ as described above,
with $U_0\to U_0\pm \sqrt{\Delta }$. We used $U_1=0.2$ and $b/a=3$ such that
the $\alpha $-value of the homogeneous system is $\sim 100$ as in Ref.\cite
{dpb}. For such large values of $\alpha $'s, the melting curves for {\em %
individual} samples display drastic multi-step behavior even for rather long
sequences of $N=10^5$; a typical example is shown in Fig.~3(a). The very
smooth melting curve for the original heterogeneous model without barrier ($%
\alpha \sim 1$) is shown in Fig.~3(b) for comparison. As in the pure PB model
with entropic barrier, we expect that the heterogeneous model 
still undergoes a second order melting transition with self-averaging melting 
curve at sufficiently large scales. We can understand 
the sharp steps in Fig.~3(a) as resulting from the first-order-like 
transition of various {\em domains}
\begin{figure}
\epsfysize=1.2in
\centerline{\epsffile{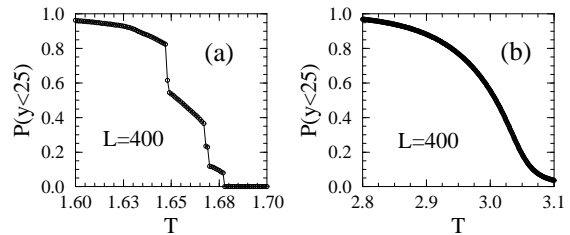}}
\vspace{\baselineskip}
\caption{ Melting curves for a {\em single} random sequence: (a)
with and (b) without the entropic barrier (see text).}
\label{F3}
\end{figure}
\noindent 
with different local transition temperatures $T_m$, shifted by variations in
the average local $U_0$ along the sequence: Due to the small width of the
transition region ($\sim 1/\alpha$), a sequence length of $O(\alpha^2)$ is
necessary just to reduce the typical shift in $T_m$ down to the size of the
transition region. Thus, the crossover length for the onset of self-averaging 
is expected to be a factor of $O(\alpha^2)$ longer for the system with
barrier. The exponential dependence of $\alpha $ on the barrier makes the
multi-step feature easily observable for realistic sequence lengths.

The domain structure are readily visualized by plotting
the full probability distribution $P_n(y)$ at various
temperatures close to the melting point (Fig.~4). 
It is seen that isolated segments of the
sequence unbind already at as much as $10\%$ below the nominal transition
temperature, indicating that the equilibrium configuration of the DNA
consists of localized bubbles of denature regions. To test whether the
experimentally observed multi-step features are indeed the result of an
effective barrier induced by the variable stiffness, one needs to determine
very accurately the form of $K(y)$.

\begin{figure}
\epsfxsize=3.2in
\centerline{\epsffile{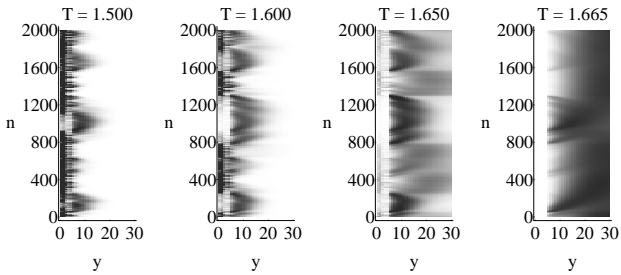}}
\vspace{\baselineskip}
\caption{ Density plot of the probability distribution $P_n(y)$ for a system
with entropic barrier as in Fig.~3(a). Darker shades correspond to larger
probability.}
\label{F4}
\end{figure}

To summarize, we have shown that variations in
base-pair binding is by itself insufficient to generate the multi-step 
melting behavior for heterogeneous DNA strands. However, 
the inclusion of a variable backbone stiffness results in an
entropic barrier which yields a sharp, first-order-like transition for the
homogeneous system, and multi-step melting for the heterogeneous system.
Asymptotically, the transition is still expected to be second order; however,
the crossover length is exponentially long.

We are grateful to helpful discussions with M.~L\"{a}ssig, H.~Li, 
and B.H.~Zimm, and especially to S.M. Bhattacharjee and S. Mukherji
for important comments.
This research is supported by a Sloan research fellowship and a
Beckman young investigator award.

\end{multicols}


\begin{references}
\bibitem{bp}  Each AT pair has two hydrogen bonds whereas each GC pair has
three hydrogen bonds.
\bibitem{sa}  As in previous studies, intra-chain contact interaction and
base-pair misalignment are neglected in this treatment.
\bibitem{transition}  Thermodynamic phase transition can occur in this 1d
system because the state variable $y_n$ in (\ref{H}) is unbounded.
\bibitem{note} For DNA, the main contribution to 
binding energy variation actually comes from the different stacking
energies gained for different complimentary base pairs~\cite{rev}.
Our model is the minimal model needed to study the effect of
binding energy variations on melting transition  --- As is usual in
critical phenomena, the details of the short-ranged function are irrelevant
in the vicinity of the phase transition.
\bibitem{bulk}  Another related problem concerning a ``defect'' in the bulk
has also been analyzed in detail recently. See Refs.~\cite{lassig1} and \cite
{bhat}.
\end{references}
\end{document}